\documentclass[prl,twocolumn,preprintnumbers,nofootinbib,tightenlines,superscriptaddress]{revtex4}

\usepackage{graphicx}
\usepackage{amssymb}
\usepackage{amsmath,mathrsfs,verbatim}
\usepackage{graphics}
\usepackage{times}
\usepackage{latexsym}

\def\msun{M_\odot}

\def\lsim{\mathrel{\raise.3ex\hbox{$<$\kern-.75em\lower1ex\hbox{$\sim$}}}}
\def\gsim{\mathrel{\raise.3ex\hbox{$>$\kern-.75em\lower1ex\hbox{$\sim$}}}}

\newcommand{\be}{\begin{equation}}
\newcommand{\ee}{\end{equation}}

\newcommand{\keV}{{\rm ~keV }}
\newcommand{\GeV}{{\rm ~GeV }}

\newcommand{\MeV}{{\rm ~MeV }}

\newcommand{\mx}{m_\chi}
\newcommand{\gx}{g_\chi}

\newcommand{\mphi}{m_\phi}

\newcounter{sec}

\begin{document}

\title{Signatures of Self-Interacting Dark Matter in the Matter Power Spectrum and the CMB}

\author{Ran Huo}
\affiliation{Department of Physics and Astronomy, University of California, Riverside, California 92521, USA}
\author{Manoj Kaplinghat}
\affiliation{Department of Physics and Astronomy, University of California, Irvine, California 92697, USA}
\author{Zhen Pan}
\affiliation{Department of Physics, University of California, Davis, California 95616, USA}
\author{Hai-Bo Yu}
\affiliation{Department of Physics and Astronomy, University of California, Riverside, California 92521, USA}

\date{\today}

\begin{abstract}

We consider a self-interacting dark matter model in which the massive dark photon mediating the self-interaction decays to light dark fermions to avoid over-closing the universe. We find that if the model is constrained to explain the dark matter halos inferred for spiral galaxies and galaxy clusters simultaneously, there is a strong indication that dark matter is produced asymmetrically in the early universe. It also implies the presence of dark radiation, late kinetic decoupling for dark matter, and a suppressed linear power spectrum due to dark acoustic damping. The Lyman-$\alpha$ forest power spectrum measurements put a strong upper limit on the damping scale and the model has little room to reduce the abundances of satellite galaxies. Future observations in the matter power spectrum and the CMB, in tandem with the impact of self-interactions in galactic halos, makes it possible to measure the gauge coupling and masses of the dark sector particles even when signals in conventional dark matter searches are absent.
\end{abstract}

\maketitle

\stepcounter{sec}
{\bf \Roman{sec}. Introduction.\;}The existence of dark matter (DM) in the universe is inferred from its gravitational influence on normal matter. Null results in terrestrial DM searches have put strong constraints on the DM interaction with the standard model particles, e.g.,~\cite{Tan:2016zwf,Akerib:2016vxi,Aaboud:2017dor}. However, it does not preclude the possibility that DM may interact strongly with itself~\cite{Spergel:1999mh,Kaplinghat:2015aga}. Strong DM self-interactions can change the inner halo structure, leading to a better agreement with small-scale observations than the cold DM (CDM) model (see~\cite{Tulin:2017ara} for a review and reference therein). In particular, kinetic thermalization due to the DM self-collisions ties the baryonic and DM distributions in galaxies together~\cite{Kaplinghat:2013xca,Elbert:2016dbb}. It has been shown~\cite{Kamada:2016euw,Creasey:2016jaq} that this can lead to the observed diversity in rotation curves of low and high surface brightness galaxies~\cite{Oman:2015xda,Karukes:2016eiz}.

A generic feature of self-interacting DM models is the existence of a light force carrier for mediating strong DM self-interactions in galactic halos. A mass hierarchy between the mediator and the DM particle is required to get a self-scattering cross section that decreases for velocities of ${\cal O}$(1000 km/s) (cluster scale)~\cite{Kaplinghat:2015aga}. This mediator must decay to avoid the over-closing the universe~\cite{Lin:2011gj,Kaplinghat:2013yxa}, unless it is (almost) massless (see, e.g.,~\cite{CyrRacine:2012fz,Cline:2013pca}). The minimal models where the mediator decays to standard model particles have been strongly constrained by DM direct detection experiments~\cite{Kaplinghat:2013yxa,DelNobile:2015uua,Kahlhoefer:2017umn}, since the DM-nucleus scattering cross section is enhanced due to the smallness of the mediator mass. In addition, the $s$-wave DM annihilation can be boosted, resulting in strong constraints from indirect detection experiments~\cite{Kaplinghat:2015gha,Bringmann:2016din,Cirelli:2016rnw}.

A simple solution is to introduce a massless particle species ($f$) in the dark sector, in addition to the DM particle ($\chi$) and the mediator ($\phi$). In the early universe, $\phi$ can be in thermal equilibrium with $f$, so that its number density becomes Boltzmann suppressed when the temperature is below its mass, avoiding the over-closure problem. Since $\phi$ is not necessary to couple to the standard model in this case, conventional DM signals can be absent. Aside from usual $\chi\textup{--}\chi$ self-scattering, $\phi$ also mediates $\chi\textup{--}f$ collisions in the early thermal bath. A tight coupling between matter and radiation in the early universe will lead to a cutoff in the linear matter power spectrum~\cite{Boehm:2000gq,Lin:2000qq,Boehm:2004th,Hooper:2007tu,Feng:2009mn,CyrRacine:2012fz,Aarssen:2012fx,Cyr-Racine:2013fsa,Buckley:2014hja,Ko:2014bka,Cherry:2014xra,Cyr-Racine:2015ihg,Binder:2016pnr}. This model was invoked previously to simultaneously flatten the density profiles of dwarfs and reduce their abundances~\cite{Aarssen:2012fx,Bringmann:2013vra,Bringmann:2016ilk}.

In this {\it Letter}, we use this model to explicitly demonstrate how astrophysical observations can pin down the particle physics parameter space. After constraining it to explain the dark matter halos inferred for dwarf galaxies and galaxy clusters, we explore the presence of the damping scale and the dark radiation using the Lyman-$\alpha$ forest, satellite counts, and CMB. In particular, we highlight two major findings.
\begin{itemize}
\item{Astrophysical data favor an asymmetric production mechanism for SIDM. When we require DM self-interactions to explain the diversity of inferred dark matter halo profiles in dwarf galaxies to clusters of galaxies, there is a minimal annihilation cross section for the inevitable process, $\chi\bar{\chi}\rightarrow\phi\phi$. For symmetric DM (both DM and anti-DM particles are equally populated), the allowed DM mass is in the narrow range of $\sim9\textup{--}240~{\rm MeV}$. The corresponding coupling constants have to be unnaturally small to give rise to a relic density consistent with the observed value.}
\item{The Lyman-$\alpha$ forest power spectrum measurements mute the impact of the damping scale (induced by the $\chi\textup{--}f$ interaction). We explicitly demonstrate that the kinetic decoupling temperature dictates the deviation of the SIDM matter power spectrum from the standard CDM case and map it to the warm DM (WDM) mass space. After taking into account the most recent Lyman-$\alpha$ constraints, we show that the model is unlikely to solve the missing satellites problem as suggested in~\cite{Aarssen:2012fx,Bringmann:2013vra,Bringmann:2016ilk}.}
\end{itemize}

\medskip%%%%%%%%%%%%%%%%%%%%%%%%%%%%%%%%%%%%%%%%%%%%%%%%%%%%%%%%%%%%%%%%%%%%%%%%%%%%%%%%%%%%%%%%%%%%%%

\stepcounter{sec}
{\bf \Roman{sec}. A Constrained Simplified SIDM Model} We consider a simplified SIDM model with the following interaction Lagrangian~\cite{Hooper:2007tu,Aarssen:2012fx},
\begin{equation}
\label{eq:theory}
{\cal L_{\rm int}}=-ig_\chi\bar{\chi}\gamma^\mu\chi\phi_\mu+m_\chi\bar{\chi}\chi+\frac{1}{2}\mphi^2\phi^\mu\phi_\mu-ig_f\bar{f}\gamma^\mu f\phi_\mu,
\end{equation}
where we assume that the SIDM particle ($\chi$) and the massless fermion $f$ couple to a gauge boson ($\phi$) with coupling constants $g_\chi$ and $g_f$, respectively. We assume $g_\chi=g_f$, since they are expected to be similar from the model building perspective. The dark sector could evolve independently from the visible sector in the early universe, and we use $\xi$ to parameterize the ratio of dark-to-visible temperatures, $T_f/T_\gamma$~\cite{Feng:2008mu,Feng:2009mn}. This model, with four parameters in total ($g_\chi, m_\chi, m_\phi, \xi$), can be regarded as a simplified version of more general and complex hidden charged DM models~\cite{Feng:2008mu,Feng:2009mn}.

\begin{figure}[t!]
\includegraphics[scale=0.65]{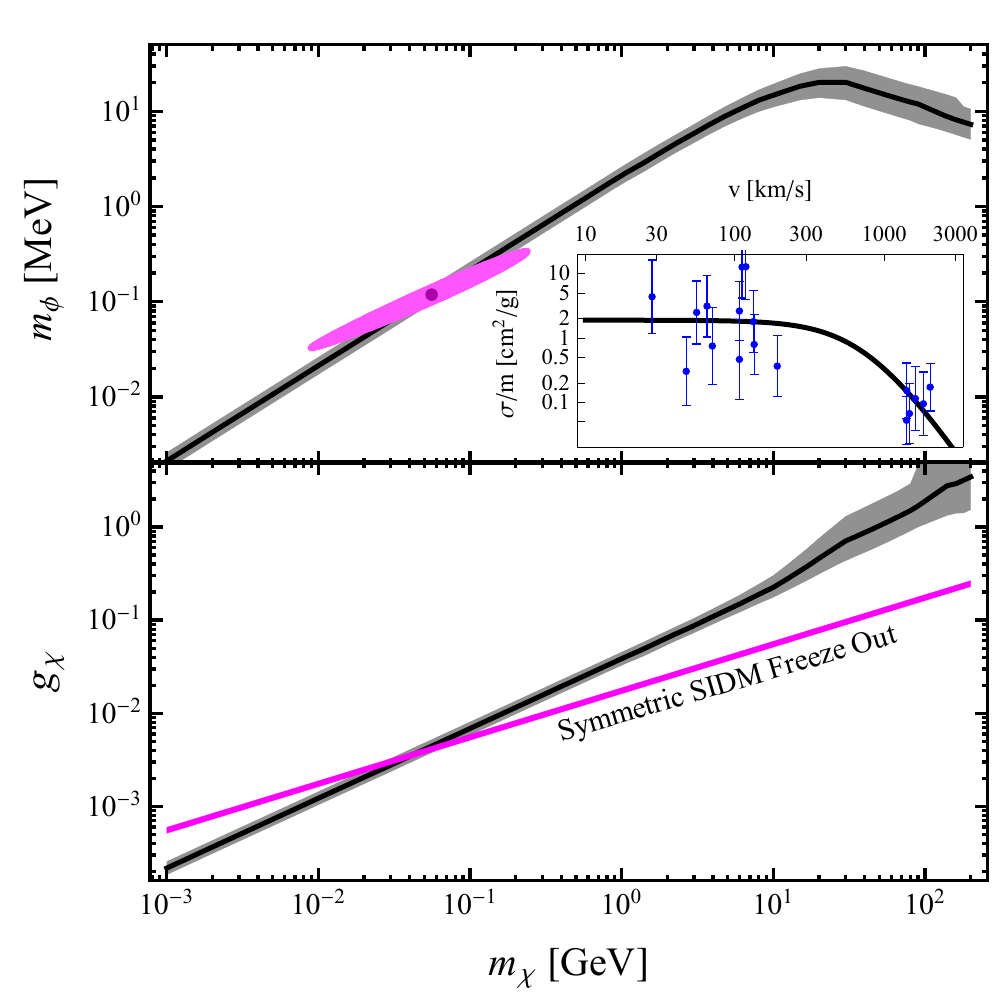}
\caption{SIDM parameter space (2$\sigma$ CL) favored by a wide range of astrophysical data from dwarf galaxies to galaxy clusters, for both asymmetric (gray) and symmetric DM (magenta). {\it Inset:} the DM self-scattering cross section vs. velocity for a best fit case and the data points with error bars are from~\cite{Kaplinghat:2015aga}.}
\label{fig:mxvsAll}
\end{figure}

In the early universe, DM particles can annihilate to the mediator. For symmetric DM, the required annihilation cross section is $\xi\times6\times10^{-26}~{\rm cm^3/s}$, which fixes $g_\chi\approx0.02(\mx/{\rm GeV})^{\frac{1}{2}}\xi^{\frac{1}{4}}$ (we take $\xi=0.48$), as denoted in Fig.~\ref{fig:mxvsAll} (lower, magenta). We further determine $\mphi$ for given $\mx$ by fitting to the preferred $\sigma_{\chi\chi}/\mx$ values in~\cite{Kaplinghat:2015aga}, which are extracted from SIDM fits to galactic rotation curves and lensing and kinematic measurements in clusters of galaxies. We find the allowed DM mass range is very limited, $9\textup{--}240~{\rm MeV}$, after {\em simultaneously} imposing $\sigma_{\chi\chi}/\mx\gtrsim1~{\rm cm^2/g}$ (galaxies) and $0.1~{\rm cm^2/g}$ (clusters), as shown in Fig.~\ref{fig:mxvsAll} (upper, magenta). Without the cluster constraint, it is possible to have strong DM self-interactions in galaxies for larger $\mx$, due to the non-perturbative enhancement effects in the quantum and classical regions~\cite{Feng:2009hw,Tulin:2013teo}, where $\sigma_{\chi\chi}/\mx$ has a strong velocity dependence and becomes negligible in clusters.

For asymmetric DM, the abundance is set up by a primordial DM asymmetry~\cite{Zurek:2013wia,Petraki:2013wwa}. We do not impose a prior constrain on $g_\chi$ from the abundance consideration and determine both $\mphi$ and $g_\chi$ from the cross section measurements, as shown in Fig.~\ref{fig:mxvsAll} (gray). Compared to the symmetric case, it is clear that asymmetric DM has a much larger mass range to be consistent observations from dwarfs to clusters. When $\mx\gtrsim200~{\rm GeV}$, the model becomes non-perturbative. For $\mx\gtrsim 40~{\rm MeV}$, $\chi\bar{\chi}\rightarrow\phi\phi$ itself can deplete the symmetric component~\cite{Zurek:2013wia}. While for a smaller mass, additional annihilation channels are required. In the rest of the paper, we will focus on asymmetric DM with $\mx\gtrsim1~{\rm GeV}$ and see {\em these constraints provide concrete predictions for the matter power spectrum and the CMB}.

The presence of light fermions $f$ contributes to the relativistic degrees of freedom parameterized as $N_{\rm eff}=3.046+\Delta N_{\rm eff}$, with $\Delta N_{\rm eff}=(11/4)^{4/3}\xi^4$. The temperature ratio, $\xi$, remains constant through kinetic decoupling and later because there is no entropy transfer in the two sectors. The analysis of Planck data indicates that $N_{\rm eff}=3.15\pm0.23$~\cite{Ade:2015xua}, which can be recast as an upper bound on the temperature ratio {\it at} the recombination epoch, $\xi\lesssim0.62$, at $2\sigma$ CL. After fitting to the Planck 2015 polarization and temperature data,~\cite{Archidiacono:2017slj} found a stronger bound $\xi\lesssim 0.48$ for imperfect fluid at $2\sigma$ CL, with some dependence on the $\chi\textup{--}f$ interaction that we neglect.

\stepcounter{sec}
{\bf \Roman{sec}. Dark Radiation, Acoustic Damping and Kinetic Decoupling.\;} In the early universe, the elastic scattering process $\chi f\rightarrow\chi f$ can damp the linear power spectrum of SIDM. When the momentum transfer rate goes below the Hubble rate, kinetic decoupling occurs. We calculate the visible sector temperature when kinetic decoupling of dark matter happens as~\cite{Aarssen:2012fx,Cyr-Racine:2015ihg}
\begin{eqnarray}
\label{eq:Tkd}
T_\text{kd}\approx\frac{1.38~{\rm keV}}{\sqrt{g_\chi g_f}}\left[\frac{\mx}{\rm 100~GeV}\right]^{\frac{1}{4}}\left[\frac{\mphi}{\rm 10~MeV}\right]\left[\frac{g_\ast}{3.38}\right]^\frac{1}{8}\left[\frac{0.5}{\xi}\right]^{\frac{3}{2}}
\end{eqnarray}
where $g_*$ is the number of massless degrees of freedom at decoupling.

\begin{figure}[t!]
\includegraphics[scale=0.5]{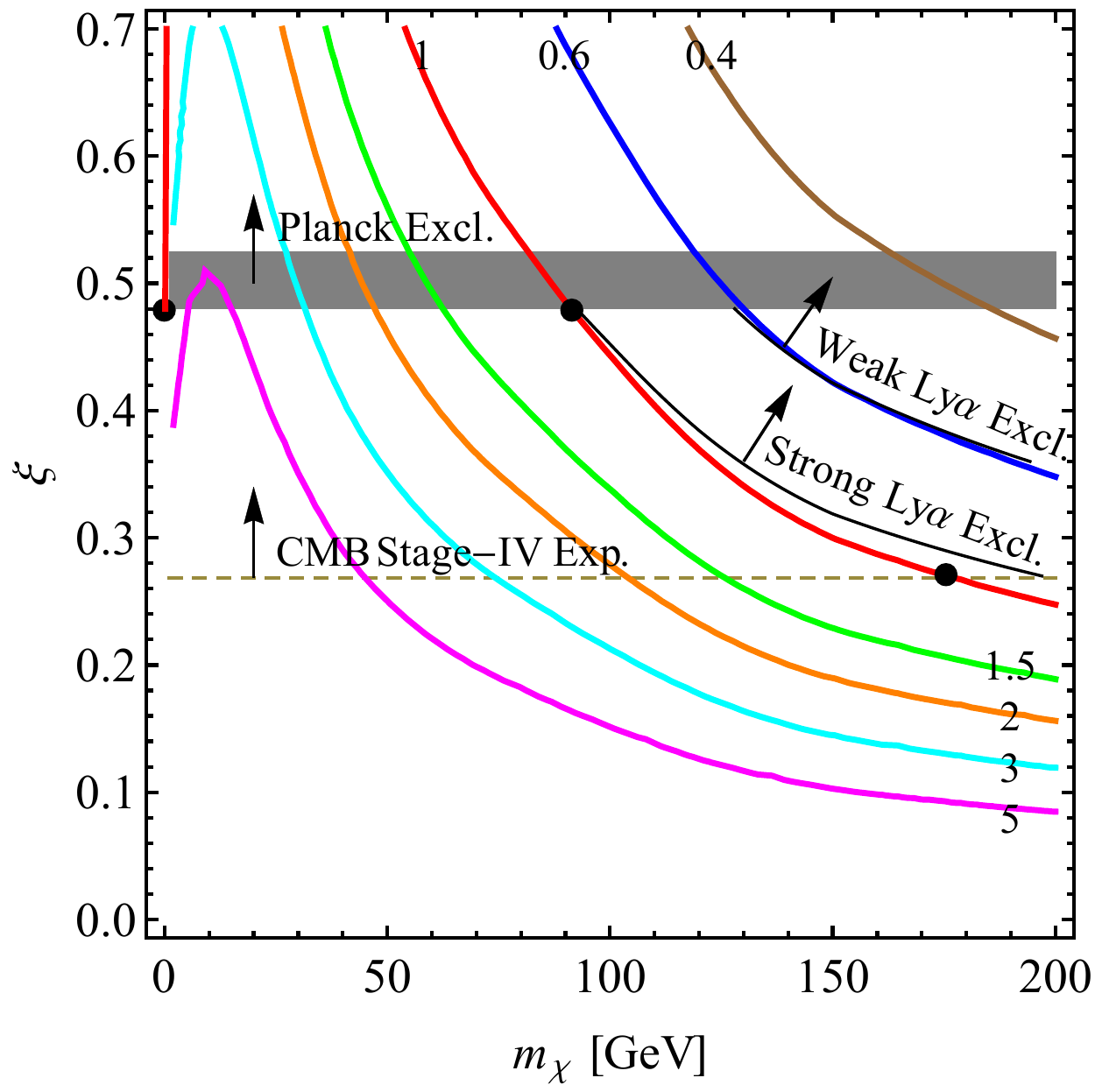}
\caption{SIDM kinetic decoupling temperature contours on the $m_\chi$ vs $\xi$ plane, $T_{\rm kd}=0.4\textup{--}5~{\rm keV}$. For given $\mx$, $\mphi$ and $\gx$ are fixed as their best fit values shown in~Fig.~\ref{fig:mxvsAll}. The regions above the arrows are constrained by the Planck~\cite{Archidiacono:2017slj} and Lyman-$\alpha$ (corresponding to the $5.3~{\rm keV}$ and $3.5~{\rm keV}$ WDM limits~\cite{Irsic:2017ixq}) observations, respectively. The horizontal line denotes the projected sensitivity of CMB Stage-IV experiments. Black dots denote the cases with their matter power spectra presented in Fig.~\ref{fig:VaryXiMass}. }
\label{fig:Tkdcontour}
\end{figure}

Fig.~\ref{fig:Tkdcontour} shows the $T_{\rm kd}$ contours for the SIDM model. There is clear degeneracy between $\mx$ and $\xi$, i.e., a stronger momentum transfer rate in the $\chi\textup{--}f$ collision can compensate a colder hidden sector thermal bath in determining $T_{\rm kd}$. In the high mass regime, $\mx\gtrsim9~{\rm GeV}$, $\mx$ increases with decreasing $\xi$ to keep a constant decoupling temperature. As shown in Fig.~\ref{fig:mxvsAll}, when $\mx$ increases from $20$ to $200\GeV$, $\mphi$ decreases from $20$ to $7\MeV$ and $\gx$ increases from $0.5$ to $3.5$.
The net result is a larger momentum transfer rate for larger $\mx$ and hence a colder hidden sector to maintain the same $T_{\rm kd}$. For $\mx\lesssim9~{\rm GeV}$, this behavior changes because the required $\mphi$ increases sharply with $\mx$, suppressing momentum transfer.

For given $\xi$, $T_{\rm kd}$ reaches its maximum, $5~{\rm keV}(0.5/\xi)^{3/2}$, when $\mx\approx9~{\rm GeV}$. If the two sectors were thermalized after inflation, e.g., through the collision process mediated by the inflaton~\cite{Adshead:2016xxj}, then the temperature ratio is $\xi \approx 0.5$ ($\Delta N_{\rm eff} \approx 0.24$) and we predict a maximal decoupling temperature of $5~{\rm keV}$,  which we use to set a lower limit on the minimum halo mass as we discuss in Sec. V. If the visible sector has additional massive new particles, $\xi$ could be lower. For example, with the minimal supersymmetric standard model, we get $\xi \approx0.43$ ($\Delta N_{\rm eff}\approx0.13$). {\em This is within the reach of the CMB-S4 experiment}, with a projected sensitivity of $\Delta N_{\rm eff}\approx0.02$ ($\xi\approx0.27$) ~\cite{Abazajian:2016yjj}.

\begin{figure}[t!]
\includegraphics[scale=0.5]{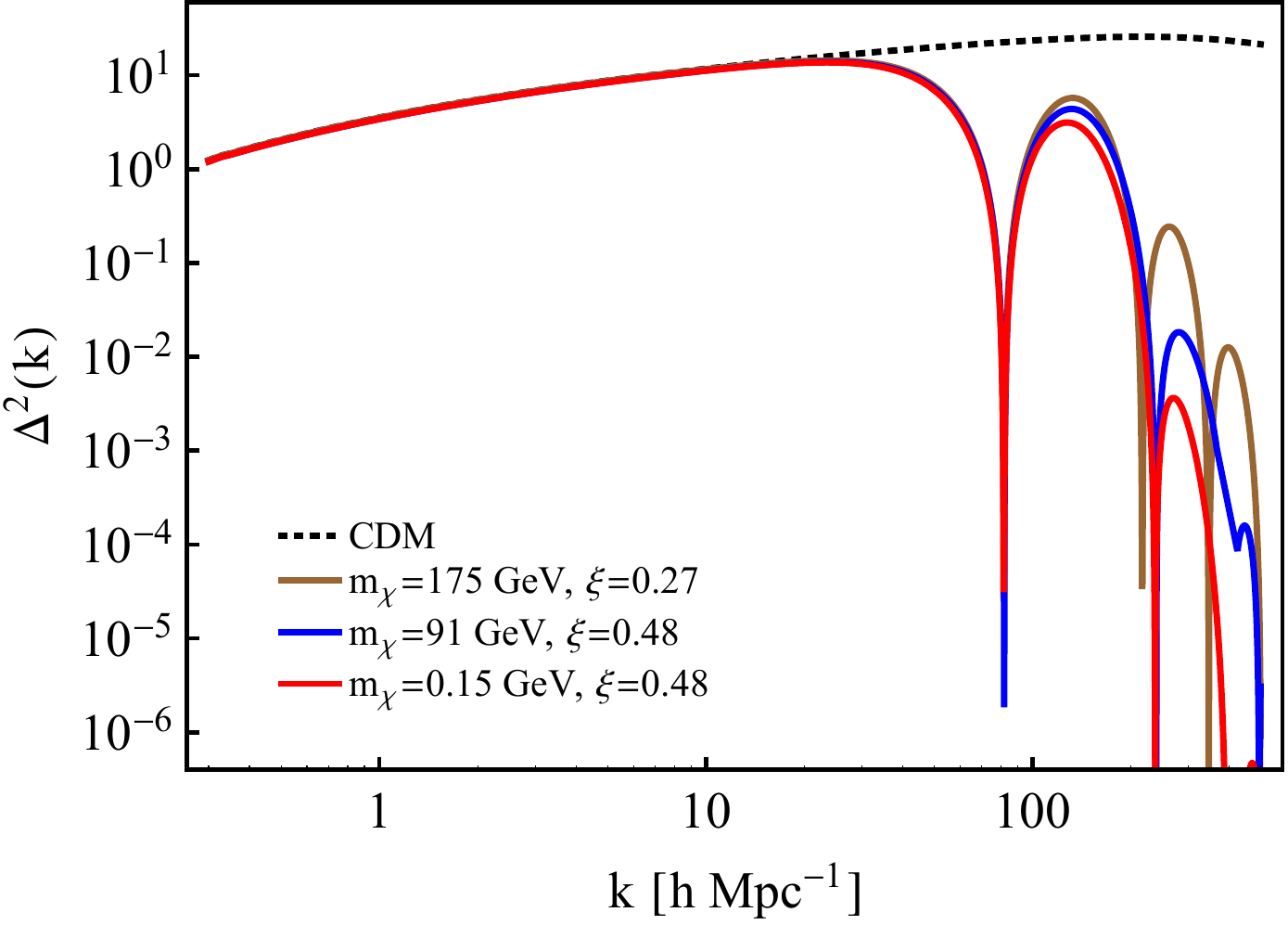}
\caption{The linear matter power spectra are similar at fixed $T_{\rm kd}$, here $1~{\rm keV}$. The models shown have parameters, $\mx=175~{\rm GeV}$ ($\xi=0.27$), $\mx=91$ and $0.15~{\rm GeV}$ ($\xi=0.48$).}
\label{fig:VaryXiMass}
\end{figure}

In Fig.~\ref{fig:VaryXiMass}, we compare the matter power spectra for three cases (denoted by the black dots in Fig.~\ref{fig:Tkdcontour}) with fixed $T_{\rm kd}$, generated using the modified version of the Boltzmann code CAMB~\cite{Lewis:2002ah} developed for the ETHOS simulations~\cite{Cyr-Racine:2015ihg}. For the model parameters, $T_{\rm kd} \ll \mx$, which implies that free-streaming effects are not relevant. The presence of dark acoustic oscillations~\cite{Loeb:2005pm,Bertschinger:2006nq,Feng:2009mn,Aarssen:2012fx,CyrRacine:2012fz} for $k \gtrsim a(T_{\rm kd})H(T_{\rm kd}) \approx 10 (T_{\rm kd}/{\rm keV}) {\rm Mpc}^{-1}$ is clearly evident in Fig.~\ref{fig:VaryXiMass}. The resulting suppression of the power spectrum is only dependent on $T_{\rm kd}$ to a good approximation, until dark Silk damping becomes important on smaller scales. The dependence on $\xi$ through the expansion rate and sound horizon is weak, which we explicitly verify in Fig.~\ref{fig:VaryXiMass} for the parameters we take.

\begin{figure}[t!]
\includegraphics[scale=0.75]{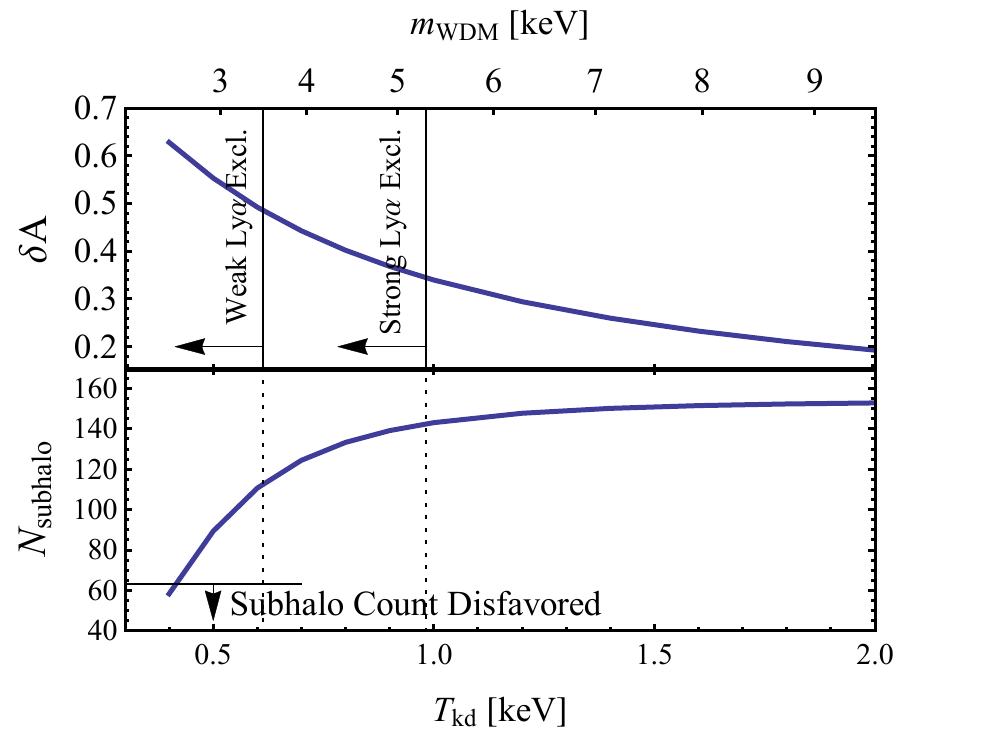}
\caption{{\em Upper}: Average deviation of one-dimensional SIDM power spectrum normalized to CDM vs. the kinetic decoupling temperature for $\xi=0.48$. We also show the corresponding thermal WDM mass that gives rise to the same $\delta A$. The strong (weak) Lyman-$\alpha$ constraints exclude thermal WDM with a mass below $5.3~{\rm keV}~(3.5~{\rm keV})$ at $95\%$ CL~\cite{Irsic:2017ixq}, which correspond to the lower limits on $T_{\rm kd}$ in the SIDM model, $T_{\rm kd}\approx1~{\rm keV}~(0.6~{\rm keV})$. {\em Lower}: The number of subhalos with masses larger than $10^8M_\odot/h$ in a MW-sized halo predicted in the SIDM model. A conservative lower limit of 63 using counts of satellites in the MW is shown by the short horizontal line. Given the strong (weak) Lyman-$\alpha$ constraints from Ref.~\cite{Irsic:2017ixq}, the number of subhalos in the SIDM model can only be suppressed by $10\%~(30\%$) compared to the CDM prediction, leaving little room for the model to impact the abundances of satellite galaxies.
}
\label{fig:WDMTkd}
\end{figure}

\stepcounter{sec}

{\bf \Roman{sec}. Lyman-$\alpha$ Constraints.\;} The Lyman-$\alpha$ forest absorption spectrum measures the neutral hydrogen density fluctuation on very large scales. Since the density of H atoms tracks that of the DM distribution, the Lyman-$\alpha$ forest can be used to constrain DM properties~\cite{Narayanan:2000tp,Viel:2005qj,Abazajian:2005xn,Viel:2013apy,Wang:2013rha,Murgia:2017lwo,Irsic:2017yje,Armengaud:2017nkf,Krall:2017xcw}. In particular, it has put strong constrains on WDM models, where the free-streaming effect damps the DM linear power spectrum. A recent combined analysis of XQ-100 and HIRES/MIKE samples put a lower limit on the thermal WDM mass, $5.3~{\rm keV}$ at $2\sigma$ CL, when the temperature evolution of the inter galactic medium is modeled as a power law in redshift~\cite{Irsic:2017ixq}. The limit is relaxed to $3.5~{\rm keV}$, if one allows a non-smooth evolution of the temperature with sudden temperature changes up to $5000~{\rm K}$~\cite{Irsic:2017ixq}.

To recast the Lyman-$\alpha$ constraints on thermal WDM as constraints on the SIDM damping scale, we use the estimator introduced in~\cite{Irsic:2017ixq}, $\delta A=(A_{\rm CDM}-A)/A_{\rm CDM}$, where $A=\int^{k_{\rm max}}_{k_{\rm min}}dk P_{\rm 1D}(k)/P^{\rm CDM}_{\rm 1D}(k)$ with $P_{\rm 1D}(k)=(1/2\pi)\int^{\infty}_k dk'k'P(k')$. $\delta A$ measures the power suppression relative to CDM. In calculating $\delta A$ for the model, we compute the 3D linear matter power spectrum, $P(k')$, at redshift $z=0$, and take $k_{\rm max}=20 h/{\rm Mpc}$ and $k_{\rm min}=0.5 h/{\rm Mpc}$ for the range of scales probed~\cite{Irsic:2017ixq}, with $h=0.67$. In Fig.~\ref{fig:WDMTkd} (upper), we map $T_{\rm kd}$ to $\delta A$ for $\xi=0.48$. In comparison, we also compute $\delta A$ for WDM and find $5.3~{\rm keV}~(3.5~{\rm keV}$) WDM and SIDM with $T_{\rm kd}\approx1~{\rm keV}~(0.6~{\rm keV})$ have the same $\delta A$. In Fig.~\ref{fig:Tkdcontour}, we also show the same constraints in the $\xi-m_\chi$ plane.

A non-zero $\Delta N_{\rm eff}$ delays matter-radiation equality and suppresses growth, which is reflected in the matter power spectrum. However, the redshift of equality is measured to roughly 1\%~\cite{Ade:2015xua}, which can be obtained by changing other cosmological parameters~\cite{Hou:2011ec}. For simplicity, we have fixed $\Lambda$CDM cosmological parameters to the Planck best-fit values~\cite{Ade:2015xua} in calculating the matter power spectrum. This is a good approximation because $\Delta N_{\rm eff}$ is small for $\xi=0.48$. Changing to $\xi=0.27$ (CMB-S4 predicted sensitivity) in our approximate analysis only weakens the constraints on $T_{\rm kd}$ by 10\%.

\begin{figure}[t!]
\includegraphics[scale=0.57]{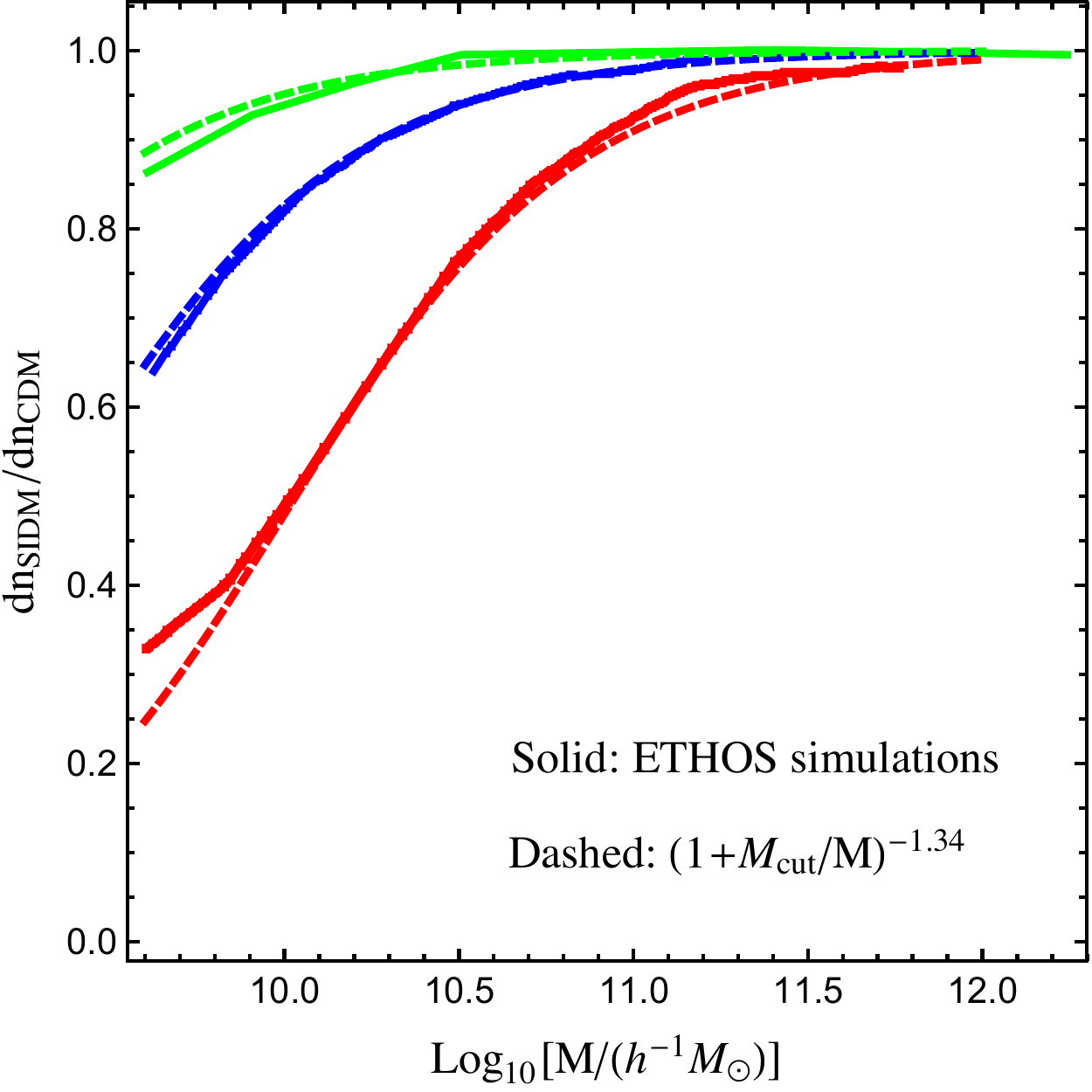}\;\;\;\;\;\;
\caption{Normalized halo mass functions for ETHOS-1 (red), 2 (blue), and 3 (green), from the simulations~\cite{Vogelsberger:2015gpr} (solid) and our analytical scaling relation (dashed), $dn_{\rm SIDM}/dn_{\rm CDM}=(1+M_{\rm cut}/M)^{-1.34}$.}
\label{fig:HMFfit}
\end{figure}
{\bf \Roman{sec}. The halo mass function.} A damped DM matter power spectrum will lead to a decrease in the number of low mass field halos and subhalos. The onset of this suppression in the field HMF is controlled by $M_\text{cut}\approx0.7\times10^8(\keV/T_\text{kd})^3~\msun$~\cite{Vogelsberger:2015gpr,Loeb:2005pm,Bertschinger:2006nq}. We find that the ansatz ${\rm d}n_{\rm SIDM}/{\rm d}M = (1+M_{\rm cut}/{M})^{-1.34} {\rm d}n_{\rm CDM}/{\rm d}M$ reproduces the field HMFs in the ETHOS simulations well, see Fig.~\ref{fig:HMFfit} for comparison.

\begin{figure*}[t!]
\includegraphics[scale=0.6]{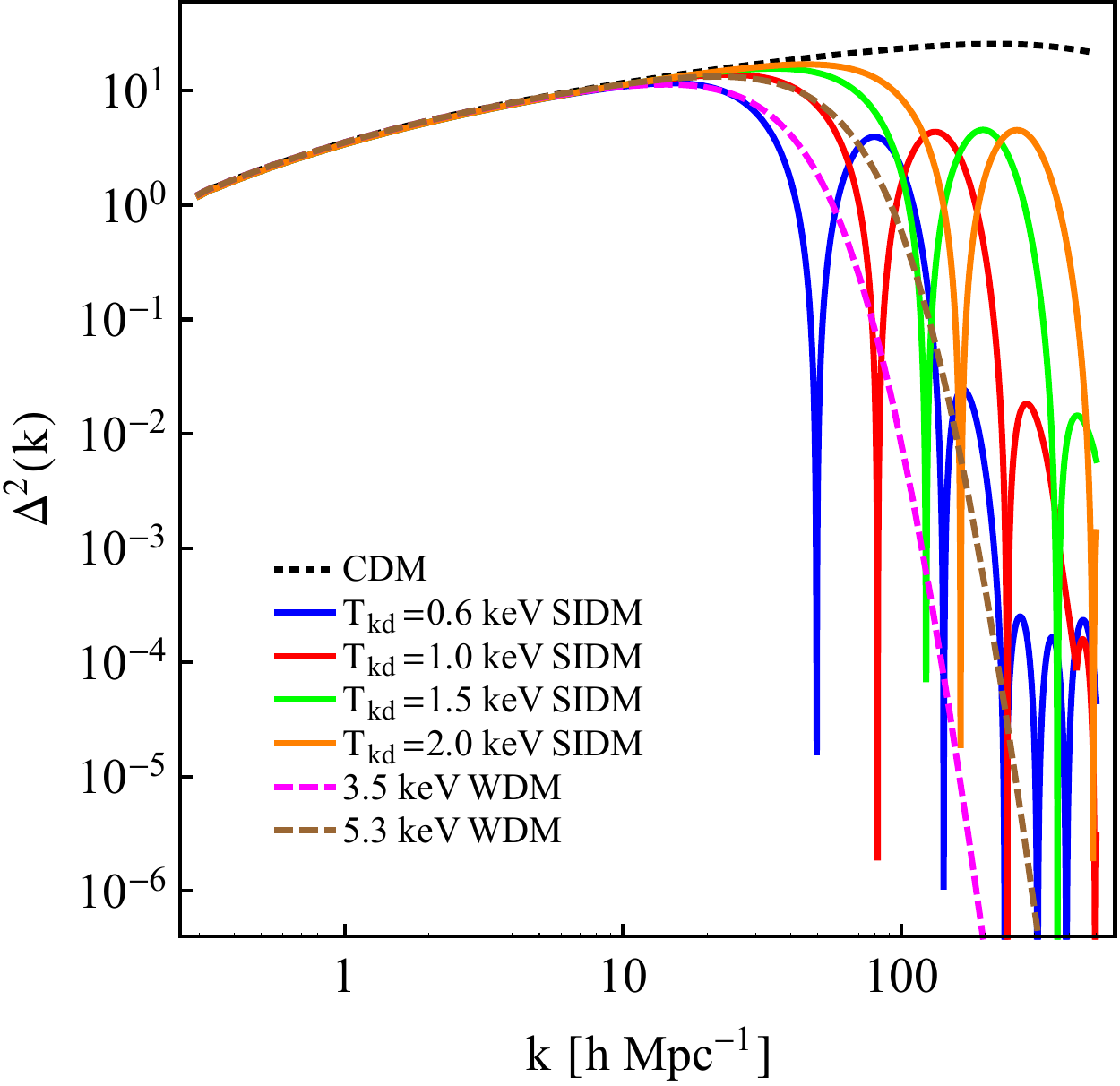}\;\;\;\;\;\;
\includegraphics[scale=0.6]{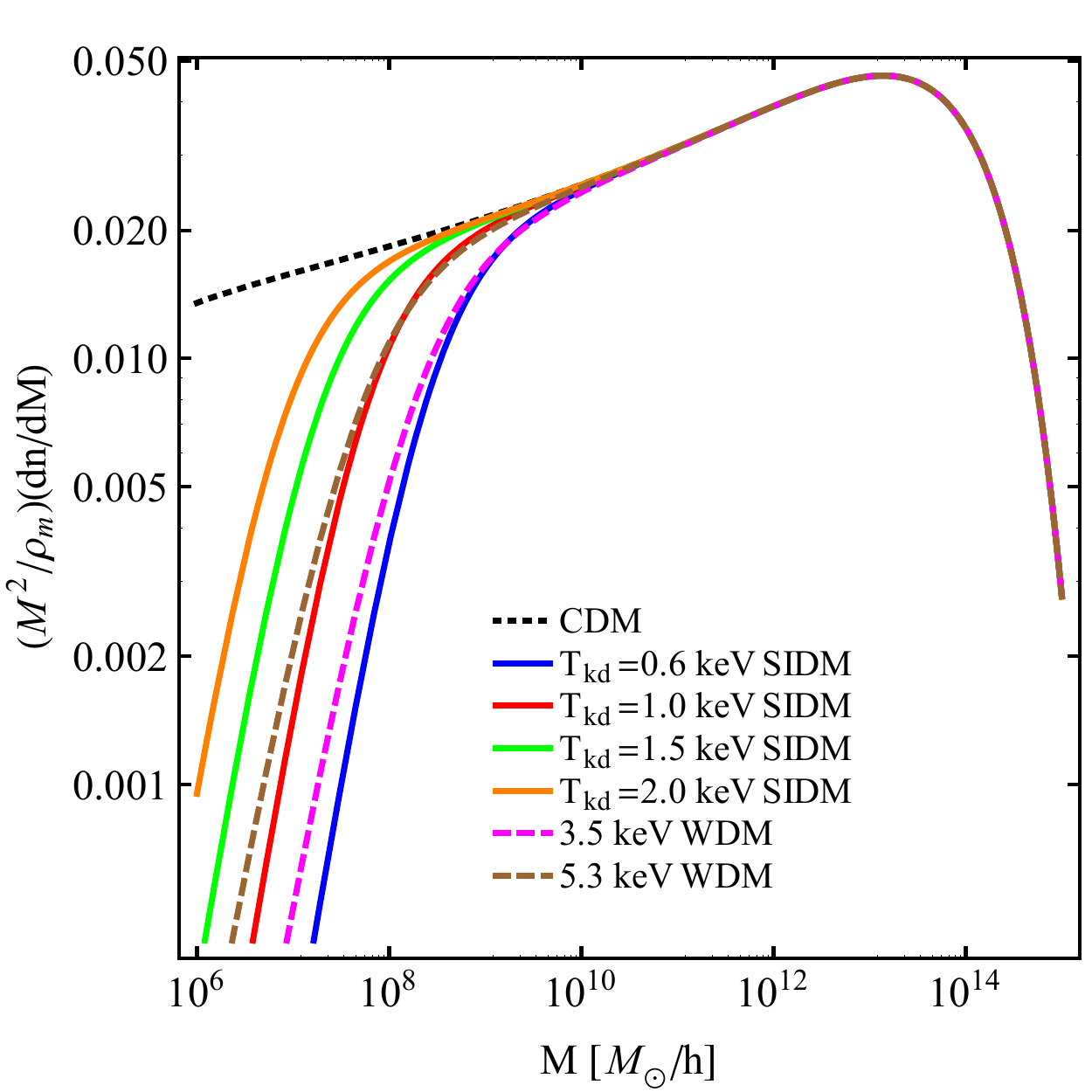}
\caption{{\em Left:} The linear matter power spectra for four SIDM benchmark cases, consistent with observations across scales from ${\rm kpc}$ to ${\rm Gpc}$. We also show $5.3\;{\rm keV}$ ($3.5\;{\rm keV}$) thermal WDM, corresponding to the strong (weak) lower bounds from the Lyman-$\alpha$ constraints derived in Ref.~\cite{Irsic:2017ixq}, and CDM. {\em Right}: Field halo mass functions, for our SIDM benchmark cases, together with thermal warm DM and CDM. We use the extended Press-Schechter theory~\cite{Bond:1990iw,Sheth:2001dp} to generate the CDM halo mass function with the fitting parameters given in~\cite{Prada:2011jf}, and multiply it with the scaling relation for the SIDM cases (see Fig.~\ref{fig:HMFfit}). The warm DM ones are estimated with the fitting formula in~\cite{Schneider:2011yu}.   }
\label{fig:HMF}
\end{figure*}
The predicted kinetic decoupling temperature of the SIDM model constrained to fit galaxy and cluster-scale halos (Fig.~\ref{fig:mxvsAll}) and allowed by Lyman-$\alpha$ constraints (Fig.~\ref{fig:WDMTkd}) is in the range of $0.6\textup{--}5~{\rm keV}$. The upper limit is obtained if the two sectors were thermalized in the early universe. The corresponding minimum halo mass is $M_{\rm cut}\sim10^8\textup{--}10^5M_{\odot}$. In Fig.~\ref{fig:HMF}, we show the power spectra for four SIDM benchmark models and two WDM models, together with their HMFs for field halos.

Satellite counts provide an important constraint on the HMF~\cite{Polisensky:2010rw,Horiuchi:2013noa}. We follow the procedure in~\cite{Schneider:2014rda} (see Eq.~17 therein) to calculate the subhalo mass function for the model. Assuming a MW halo mass $1.7\times10^{12}M_{\odot}/h$, we obtain the number of subhalos with masses larger than $10^8M_{\odot}/h$, $N_{\rm subhalo}$, for given $T_{\rm kd}$, as shown in Fig.~\ref{fig:WDMTkd} (lower). We demand that this number should be greater than $63$, which is the sum of $11$ classical satellites and $15$ SDSS satellites $\times$ 3.5 for incomplete sky coverage~\cite{Polisensky:2010rw,Schneider:2016uqi}. This constraint is weaker than the Lyman-$\alpha$ limits but we expect it to be a comparable constraint as more satellites are discovered.

With the new satellites in the DES footprint~\cite{Koposov:2015cua,Bechtol:2015cbp,Drlica-Wagner:2015ufc,Fermi-LAT:2016uux}, the total number of satellites (corrected for selection biases) may be consistent with $\Lambda$CDM expectations~\cite{Tollerud:2008ze,Hargis:2014kaa,Dooley:2016xkj}. This is still an open issue. The potential mismatch between the number of predicted subhalos and the observed satellites (``missing satellites problem''~\cite{Klypin:1999uc,Moore:1999nt}) has been used to motivate the presence of dark acoustic damping~\cite{Aarssen:2012fx}. Our analysis shows that the constraints from the Lyman-$\alpha$ forest power spectrum, assuming no significant unmodeled systematic effects, leave little room to modify the abundance of satellites. For the strong (weak) Lyman-$\alpha$ constraint, $T_{\rm kd}=1~{\rm keV}~(0.6~{\rm keV})$, the corresponding number of subhalos for masses larger than $10^8M_{\odot}/h$ is $N_{\rm subhalo}\approx142~(113)$, as shown in Fig.~\ref{fig:WDMTkd} (lower), which is only $10\%~(30\%)$ less than that predicted in the CDM model. Since $T_{\rm kd}$ determines the damped power spectrum, modifying the particle masses or couplings will not change this result.

On the other hand, self-interactions (leading to core formation) could change the distribution of satellites in the inner region of MW and Andromeda due to tidal effects~\cite{Penarrubia:2010jk}, while the distribution in the outer parts remains similar to the collisionless case~\cite{Vogelsberger:2012ku,Rocha:2012jg}. It is also possible that the early star formation feedback effects are different in cored SIDM halos and this may impact the faint-end luminosity function. These effects are clearly relevant for the ``missing satellites problem,'' and remained to be quantified. Observationally, we expect LSST to provide a definite statement in terms of the census of the ultra-faint satellites~\cite{Hargis:2014kaa}. In addition, the dark subhalos (or ultra-faint galaxies) could be discovered gravitationally through lensing~\cite{Mao:1997ek,Chiba:2001wk,Dalal:2001fq,Kochanek:2003zc,Vegetti:2012mc,Asadi:2015nyw,Inoue:2015lma,Hezaveh:2016ltk}, and tidal streams in the MW~\cite{Johnston:1997fv,Ibata:2001iv,Johnston:2001wh,SiegalGaskins:2007kq,Yoon:2010iy,Carlberg:2012ur, Erkal:2015kqa,Bovy:2016irg}.

{\em Does the model solve the too-big-to-fail problem at the bright end of the satellite luminosity function}~\cite{BoylanKolchin:2011dk,BoylanKolchin:2011de}? The ETHOS-4 model has $T_{\rm kd}=0.5~{\rm keV}$ and it agrees broadly with observations of the dwarf spheroidals in the MW~\cite{Vogelsberger:2015gpr}. This is mainly due to damping (see also~\cite{Schewtschenko:2015rno}) because $\sigma_{\chi\chi}/\mx\sim0.2~{\rm cm^2/g}$ in this model, which cannot  change the inner halo of dwarf galaxies significantly~\cite{Zavala:2012us}. Given the Lyman-$\alpha$ constraint, the small-scale power in our case is less suppressed ($T_{\rm kd}\gtrsim1~{\rm keV}~(0.6~{\rm keV})$) compared to ETHOS-4 model. However, in our case, the self-interaction cross section is large in dwarf galaxies ($\sigma_{\chi\chi}/\mx\sim2~{\rm cm^2/s}$), which would lower the subhalo densities due to core formation to roughly the right values~\cite{Elbert:2014bma,Vogelsberger:2012ku}. It is of interest to perform a more detailed assessment of the too-big-to-fail problem in our case.

In the future, if a cut-off scale in the HMF is observed, we can derive its corresponding $T_{\rm kd}$ and fix the relation in the $\xi\textup{--}\mx$ plane (see Fig.~\ref{fig:Tkdcontour}). The particle parameters, e.g., $\mx$, $\mphi$, and $\gx$, can be further determined (up to the two-fold degeneracy) if the presence of dark radiation is detected or even a stronger upper bound on $\xi$ is obtained. Similar conclusions seem to also apply to the case of the atomic dark matter model~\cite{Kaplan:2009de,CyrRacine:2012fz,Cline:2013pca,Buckley:2014hja}, when it is constrained to solve the small-scale puzzles~\cite{Boddy:2016bbu}.

\stepcounter{sec}
{\bf \Roman{sec}. Conclusions.\;} SIDM is a compelling alternative to CDM. It keeps all the success of CDM on large scales, while modifying the inner halo structure in accord with observations. Using a simplified particle physics realization, we have shown that SIDM generically prefers asymmetric DM, and predicts the existence of dark radiation and a damped linear DM power spectrum, with the damping scale set by the self-scattering cross section and the temperature ratio between the two sectors. We have mapped out the favored model parameters, combining observations of stellar kinematics of spiral galaxies and galaxy clusters, the CMB, and Lyman-$\alpha$ forest, to narrow down the SIDM model parameter space. The predictions from this viable region of parameter space are the presence of dark radiation and the cut-off in the mass function of halos, both potentially observable in the future.

{\it Acknowledgments}: We thank Anson D'Aloisio for useful discussion. This work was supported by the National Science Foundation Grant PHY-1620638 (MK), UC Davis Dissertation Year Fellowship (ZP), and the U.~S.~Department of Energy under Grant No.~DE-SC0008541 (HBY). HBY acknowledges support from the Hellman Fellows Fund.

\bibliography{DMstructure}

\end{document}